\documentstyle[12pt]{article}
\advance\hoffset by -7mm
 \makeatletter
\@addtoreset{equation}{section} \makeatother

\renewcommand{\title}[1]{\null\vspace{25mm}

\noindent{\Large{\bf #1}}\vspace{10mm} }
\newcommand{\authors}[1]{\noindent{\large #1}\vspace{3mm}

}
\newcommand{\address}[1]{\noindent #1\vspace{5mm}

}

\renewcommand{\abstract}[1]{\vspace{19mm}

\noindent{\small{\em Abstract.} #1}\vspace{2mm} }

\begin{document}
\title{The Harmonic Oscillator with Dissipation\\[2mm]
within the Theory of Open Quantum Systems}

\authors{A. Isar}
\address{Department of Theoretical Physics, Institute of Atomic Physics\\
Bucharest-Magurele, POB MG-6, Romania\\
Internet: isar@theory.nipne.ro}
\abstract{ Time evolution of the
expectation values of various dynamical operators of the harmonic
oscillator with dissipation is analitically obtained within the
framework of the Lindblad theory for open quantum systems. We deduce
the density matrix of the damped harmonic oscillator from the
solution of the Fokker-Planck equation for the coherent state
representation, obtained from the master equation for the density
operator. The Fokker-Planck equation for the Wigner distribution
function, subject to either the Gaussian type or the
$\delta$-function type of initial conditions, is also solved by
using the Wang-Uhlenbeck method. The obtained Wigner functions are
two-dimensional Gaussians with different widths. }

\section{Introduction}

In the last two decades, the problem of dissipation in quantum
mechanics, i.e. the consistent description of open quantum systems,
was investigated by various authors [1-6]. Because dissipative
processes imply irreversibility and, therefore, a preferred
direction in time, it is generally thought that quantum dynamical
semigroups are the basic tools to introduce dissipation in quantum
mechanics. The most general form of the generators of such
semigroups was given by Lindblad [7-9]. This formalism has been
studied for the case of damped harmonic oscillators \cite{l2,s1,s2}
and applied to various physical phenomena, for instance, the damping
of collective modes in deep inelastic collisions in nuclear physics
\cite{p,i1} and the interaction of a two-level atom with the
electromagnetic field \cite{s3}. Recently \cite{g}, a family of
master equations, constructed in the form of Lindblad generators,
was proposed for local ohmic quantum dissipation.

This paper, dealing with the damping of the harmonic oscillator
within the Lindblad theory for open quantum systems, is concerned
with the time evolution of various dynamical operators involved in
the master and Fokker-Planck equations, in particular with the time
development of the density matrix. In \cite{i2} the Lindblad master
equation was transformed into Fokker-Planck equations for
quasiprobability distributions and a comparative study was made for
the Glauber $P$, antinormal ordering $Q$ and Wigner $W$
representations. In \cite{i3} the density matrix of the damped
harmonic oscillator was  represented by a generating function. We
shall explore the physical aspects of the Fokker-Planck equation
which is the $c$-number equivalent equation to the master equation
for the density operator. Generally the master equation gains
considerably in clarity if it is represented in terms of the Wigner
distribution function which satisfies the Fokker-Planck equation. It
is worth mentioning that these master and Fokker-Planck equations
agree in form with the corresponding equations formulated in quantum
optics [18-25].

The content of the paper is arranged as follows. In Sec.2 we review
the derivation of the master equation of the harmonic oscillator. In
order to get an insight into physical meanings of this equation, we
first split it up into several equations satisfied by the
expectation values of dynamical operators involved in the master
equation. These equations are then solved analytically. In Sec.3 we
transform the master equation into the Fokker-Planck equation by
means of the well-known methods [4,26-30]. We extract the density
matrix with the help of the solution of the Fokker-Planck equation
for the coherent state representation. Then the Fokker-Planck
equation for the Wigner distribution, subject to either the Gaussian
type or the $\delta$-function type of initial conditions, is solved
by the Wang-Uhlenbeck method. Finally, conclusions are given in
Sec.4.

\section{Master equation for the damped harmonic oscillator}

The rigorous formulation for introducing the dissipation into a
quantum mechanical system is that of quantum dynamical semigroups
\cite{d,s,l1}. According to the axiomatic theory of Lindblad
\cite{l1,l3}, the
usual von Neumann-Liouville equation ruling the time evolution of
closed quantum systems is replaced in the case of open systems by the
following equation for the density operator $\rho$:
$${d\Phi_{t}(\rho)\over dt}=L(\Phi_{t}(\rho)).\eqno (2.1)$$
Here, $\Phi_{t}$ denotes the dynamical semigroup describing the
irreversible time evolution of the open system in the Schr\"odinger
representation and $L$ the infinitesimal generator of the dynamical
semigroup $\Phi_t.$ Using the structural theorem of Lindblad \cite{l1}
which
gives the most general form of the bounded, completely dissipative
Liouville operator $L$, we obtain the explicit form of the most
general time-homogeneous quantum mechanical Markovian master equation:
$${d\rho(t)\over dt}=L(\rho(t))=-{i\over\hbar}[H,\rho(t)]+{1\over 2
\hbar}\sum_{j}([V_{j}\rho(t),V_{j}^\dagger ]+[V_{j},\rho(t)V_{j}^\dagger ]).
\eqno (2.2)$$
Here $H$ is the Hamiltonian of the system. The operators $V_{j}$ and
$V_{j}^\dagger $ are bounded operators on the Hilbert space of the
Hamiltonian.

We should like to mention that the Markovian master equations found in
the literature are of this form after some rearrangement of terms,
even for unbounded Liouville operators. In this connection we assume
that the general form of the master equation given by (2.2) is also
valid for unbounded Liouville operators.

In this paper we impose a simple condition to the operators $H,V_{j},
V_{j}^\dagger $ that they are functions of the basic observables $q$ and $p$
of the one-dimensional quantum mechanical system of such kind that the
obtained model is exactly solvable. This condition implies \cite{l2}
that
$V_{j}$ are at most first degree polynomials in $p$ and $q$ and $H$ is
at most a second degree polynomial in $p$ and $q$.
Because in the linear space of the first degree polynomials in $p$ and $q$
the operators $p$ and $q$ give a basis, there exist only two ${\bf C}$-linear
independent operators $V_{1},V_{2}$ which can be written in the form
$$V_{j}=a_{j}p+b_{j}q,~j=1,2, \eqno(2.3)$$
with $a_{j},b_{j}$ complex numbers \cite{l2}. The constant term is omitted
because its contribution to the generator $L$ is equivalent to terms in $H$
linear in $p$ and $q$ which for simplicity are assumed to be zero.
Then the harmonic
oscillator Hamiltonian $H$ is chosen of the form
$$H=H_{0}+{\mu \over 2}(pq+qp),~~~H_{0}={1\over 2m}p^2+{m\omega^2\over
2}q^2.\eqno (2.4)$$
With these choices and introducing the annihilation and creation
operators via the relations
$$q=\sqrt{{\hbar\over 2m\omega}}(a^\dagger +a), ~~p=i\sqrt{{\hbar m\omega
\over 2}}(a^\dagger -a),\eqno(2.5)$$
we have $H_{0}=\hbar\omega(a^\dagger a+1/2)$ and the Markovian master
equation takes the form
$${d\rho\over dt}={1\over 2}(D_{1}+\mu)(a^\dagger  a^\dagger \rho-a^\dagger
\rho a^\dagger )+{1
\over 2}(D_{1}-\mu)(\rho a^\dagger  a^\dagger -a^\dagger \rho a^\dagger )$$
$$+{1\over2}(D_{2}+\lambda+i\omega)(a\rho a^\dagger -a^\dagger  a\rho)+{1
\over 2}
(D_{2}-\lambda-i\omega)(a^\dagger \rho a-\rho aa^\dagger )+{\rm H.c.},\eqno
(2.6)$$
where
$$D_{1}\equiv {1\over\hbar}(m\omega D_{qq}-{D_{pp}\over m\omega}+2i
D_{pq}), ~~D_{2}\equiv {1\over\hbar}(m\omega D_{qq}+{D_{pp}\over m
\omega}). \eqno (2.7)$$
Here we used the notations:
$$D_{qq}={\hbar\over 2}\sum_{j=1,2}{\vert a_{j}\vert}^2,
~~D_{pp}={\hbar\over 2}\sum_{j=1,2}{\vert b_{j}\vert}^2,
~~D_{pq}=D_{qp}=-{\hbar\over 2}{\rm Re}\sum_{j=1,2}a_{j}^*b_{j},
~~\lambda=-{\rm Im}\sum_{j=1,2}a_{j}^*b_{j}, \eqno(2.8)$$
where $D_{pp},D_{qq}$ and $D_{pq}$ are the diffusion coefficients and
$\lambda$ the friction constant. They satisfy the following
fundamental constraints \cite{s1}:
$${\rm i})~D_{pp}>0,~~{\rm ii})~D_{qq}>0,~~{\rm iii})~D_{pp}D_{qq}-
D_{pq}^2\ge{\lambda}^2{\hbar}^2/4.\eqno(2.9)$$
In the particular case when the asymptotic state is a Gibbs state
$$\rho_G(\infty)=e^{-{H_0\over kT}}/{\rm Tr}e^{-{H_0\over kT}},
\eqno(2.10)$$
these coefficients reduce to
$$D_{pp}={\lambda+\mu\over 2}\hbar m\omega\coth{\hbar\omega\over 2kT},
~~D_{qq}={\lambda-\mu\over 2}{\hbar\over m\omega}\coth{\hbar\omega
\over 2kT}, ~~D_{pq}=0,\eqno(2.11)$$
where $T$ is the temperature of the thermal bath.

In the literature, master equations of the type (2.6) are
encountered in concrete theoretical models for the description of
different physical phenomena in quantum optics [18-24], in
treatments of the damping of collective modes in deep inelastic
collisions of heavy ions [31-34] or in the quantum mechanical
description of the dissipation for the one-dimensional harmonic
oscillator \cite{de,l,gr,h2}. A classification of these equations,
whether they satisfy or not the fundamental constraints (2.9), was
given in \cite{i3}.

The meaning of the master equation becomes clear when we transform it
into equations satisfied by various expectation values of operators
involved in the master equation, $<A>={\rm Tr}[\rho(t)A],$ where $A$
is an operator composed of the creation and annihilation operators.
Multiplying both sides of (2.6) by $a$ and taking throughout the
trace, we get
$${d\over dt}<a>=-(\lambda+i\omega)<a>+\mu<a^\dagger  >.\eqno(2.12)$$
Similarly, the equation for $<a^\dagger  >$ is given by
$${d\over dt}<a^\dagger >=-(\lambda-i\omega)<a^\dagger >+\mu<a>.\eqno(2.13)$$
In the absence of the second term on the right-hand side of (2.12) and
(2.13), the two equations represent independently a simple equation of
oscillation with damping. By coupling (2.12) to (2.13) we get a second
order differential equation for either $<a>$ or $<a^\dagger >$. For example,
we obtain:
$${d^2\over dt^2}<a>+2\lambda{d\over dt}<a>+(\lambda^2+\omega^2-\mu^2)
<a>=0,\eqno(2.14)$$
which is the equation of motion for Brownian motion of a classical
oscillator, but without the term corresponding to random process.
Because of the vanishing terms on the right-hand side, we may equally
state that (2.14) is the equation of motion with zero expectation
value of the random process. In the study of the Brownian motion of a
classical oscillator, one replaces the second-order differential
equation of motion of the type (2.14) by two equivalent first-order
differential equations \cite{w} which are precisely the Langevin
equations. Accordingly, we may say that (2.12) and (2.13) are the
Langevin equations corresponding to (2.14), but without the random
process term. The integration of (2.14) is straightforward. There are
two cases: a)$\mu>\omega$ (overdamped) and b)$\mu<\omega$
(underdamped). In the case a) with the notation $\nu^2\equiv\mu^2-
\omega^2,$ we obtain:
$$<a(t)>= e^{-\lambda t}[<a(0)>(\cosh\nu t-i{\omega\over\nu}\sinh\nu
t)+{\mu\over\nu}<a^\dagger (0)>\sinh\nu t].\eqno(2.15 {\rm a})$$
In the case b) with the notation $\Omega^2\equiv\omega^2-\mu^2$, we
obtain:
$$<a(t)>=e^{-\lambda t}[<a(0)>(\cos\Omega t-i{\omega\over\Omega}\sin
\Omega t)+{\mu\over\Omega}<a^\dagger (0)>\sin\Omega t].\eqno(2.15 {\rm b})$$
The expression for $<a^\dagger (t)>$ can be obtained simply by taking the
complex conjugate of the right-hand side of (2.15).

For the computation of quantal fluctuations of the coordinate and
momentum of the harmonic oscillator, we need the expectation values of
quadratic operators, such as $a^{+2}, a^2$ or $a^\dagger a$. The dynamical
behaviour of these operators can be well surveyed by deriving the
equations satisfied by their expectation values. By following the same
procedure as before, employed in the derivation of (2.12), we find:
$${d\over dt}<a^2>+2(\lambda+i\omega)<a^2>=2\mu<a^\dagger a>+D_1+\mu,
\eqno(2.16)$$
$${d\over dt}<a^{+2}>+2(\lambda-i\omega)<a^{+2}>=2\mu<a^\dagger a>+D_1^*+\mu,
\eqno(2.17)$$
$${d\over dt}<a^\dagger a>+2\lambda<a^\dagger a>=\mu(<a^{+2}>+<a^2>)+D_2-
\lambda.
\eqno(2.18)$$
The solutions of these equations are readily obtained by transforming
them into two differential equations satisfied by the sum and the
difference of two quadratic operators, $<a^2>$ and $<a^{+2}>$. We
obtain:
$${1\over 4}{d^2\over dt^2}(<a^2>+<a^{+2}>)+\lambda{d\over dt}(<a^2>+
<a^{+2}>)+(\lambda^2+\omega^2-\mu^2)(<a^2>+<a^{+2}>)$$
$$={1\over\hbar}[(\lambda+\mu)m\omega D_{qq}-(\lambda-\mu){D_{pp}\over
m\omega}+2\omega D_{pq}]\equiv D,\eqno(2.19)$$
$${1\over 2}{d\over dt}(<a^2>-<a^{+2}>)+\lambda(<a^2>-<a^{+2}>)+i
\omega(<a^2>+<a^{+2}>)=2i{D_{pq}\over\hbar}.\eqno(2.20)$$
The solution of (2.19) is straightforward and with the help of which
both equations (2.18) and (2.20) can be immediately solved. We find:
$$<a^2>=e^{-2\lambda t}[(1-i{\omega\over\nu})C_1 e^{2\nu t}+(1+i{
\omega\over\nu})C_2 e^{-2\nu t}-i{\mu\over\omega}C_3]+{D(\lambda-i
\omega)\over 2\lambda(\lambda^2-\nu^2)}+i{D_{pq}\over\hbar\lambda},
\eqno(2.21 {\rm a})$$
$$<a^\dagger a>=e^{-2\lambda t}[{\mu\over\nu}(C_1 e^{2\nu t}-C_2 e^{-2\nu t})
+C_3]+{1\over 2\lambda}({D\mu\over \lambda^2-\nu^2}+D_2-\lambda),
\eqno(2.22 {\rm a})$$
for the overdamped case $\mu>\omega$ and
$$<a^2>=e^{-2\lambda t}[(C_1+iC_2{\omega\over\Omega})\cos 2\Omega t+
(C_2-iC_1{\omega\over\Omega})\sin 2\Omega t-i{\mu\over\omega}C_3]+{D(
\lambda-i\omega)\over 2\lambda(\lambda^2+\Omega^2)}+i{D_{pq}\over
\hbar\lambda},\eqno(2.21 {\rm b})$$
$$<a^\dagger a>=e^{-2\lambda t}[{\mu\over\Omega}(C_1\sin 2\Omega t-C_2\cos 2
\Omega t)+C_3]+{1\over 2\lambda}({D\mu\over \lambda^2+\Omega^2}+D_2-
\lambda),\eqno (2.22 {\rm b})$$
for the underdamped case $\omega>\mu$. The expression for $<a^{+2}>$
can be obtained by taking the complex conjugates of (2.21{\rm a}) and
(2.21{\rm b}). Here, $C_1,C_2,C_3$ are the integral constants
depending on the initial expectation values of the operators under
consideration. In particular, if $D_{qq}=D_{pq}=0$ and $\mu=\lambda$
we obtain for the underdamped case $\omega>\mu$ the equations written
by Jang \cite{j} for the model on nuclear dynamics based on the second
RPA
at finite temperature. For time $t\to\infty,$ we see from (2.22) that
$$<a^\dagger a>={1\over 2\lambda}({D\mu\over \lambda^2+\omega^2-\mu^2}+D_2-
\lambda).\eqno(2.23)$$
In the particular case when the asymptotic state is a Gibbs state
(2.10), we get
$$<a^\dagger a>={1\over 2}(\coth{\hbar\omega\over 2kT}-1)=(\exp{\hbar\omega
\over kT}-1)^{-1}\equiv <n>,\eqno(2.24)$$
which is the Bose distribution. This means that the expectation value
of the number operator goes to the average thermal-phonon number at
infinity of time. From the identity
$$<a^\dagger a>=\sum_{m=0}^\infty m<m\vert\rho(t)\vert m>\eqno(2.25)$$
it follows
$$<m\vert\rho(\infty)\vert m>={<n>^m\over (1+<n>)^{m+1}}.\eqno(2.26)$$
In deriving this formula, we have made use of the identity $\sum_{m=0}
^\infty mx^m=x/(1-x)^2.$ The expression (2.26) shows that in the
considered particular case the density matrix reaches its thermal
equilibrium -- the Bose-Einstein distribution, whatever the initial
distribution of the density matrix may be. When the initial density
matrix $<m\vert\rho(0)\vert m>$ is represented by a distribution of
the form $N^m/(1+N)^{m+1}$, where $N$ stands for the average phonon
number, the relation (2.25) implies that $<a^\dagger a>=N.$ When the initial
density matrix is characterized by a distribution of the form
$${1\over m!}N^m e^{-N},\eqno(2.27)$$
(2.25) implies that $<a^\dagger a>$ becomes also $N.$ Eq. (2.27) is nothing
but a Poisson distribution. If the initial density matrix is
represented by a Kronecker delta $\delta_{ms},$ we see from (2.25)
that $<a^\dagger a>=s,$ which corresponds to the initial $s$-phonon state.

The physical observables of the harmonic oscillator can be obtained
from the expectation values of polynomials of the annihilation and
creation operators. So, for the position and momentum operators $q$
and $p$ via the relations (2.5), we can evaluate either the second
moments or variances (fluctuations), by making use of the results
(2.15), (2.21), (2.22).

\section{Fokker-Planck equations}

One useful way to study the consequences of the master equation
(2.6) for the density operator of the one-dimensional damped
harmonic oscillator is to transform it into more familiar forms,
such as the equations for the $c$-number quasiprobability
distributions Glauber $P$, antinormal ordering $Q$ and Wigner $W$
associated with the density operator \cite{i2}. In this case the
resulting differential equations of the Fokker-Planck type for the
distribution functions can be solved by standard methods [26-29]
employed in quantum optics and observables directly calculated as
correlations of these distribution functions.

\subsection{Calculation of the density matrix from the Fokker-Planck
equation}

The Fokker-Planck equation, obtained from the master equation and
satisfied by the Wigner distribution function $W(\alpha,\alpha^*,t)$,
where $\alpha$ is a complex variable, has the form \cite{i2}:
$${\partial W(\alpha,\alpha^*,t)\over\partial t}=-\{{\partial\over
\partial\alpha}[-(\lambda+i\omega)\alpha+\mu\alpha^*]+{\partial\over
\partial\alpha^*}[-(\lambda-i\omega)\alpha^*+\mu\alpha]\}W(\alpha,
\alpha^*,t)$$
$$+{1\over 2}(D_1{\partial^2\over\partial\alpha^2}+D_1^*{\partial^2
\over\partial\alpha^{*2}}+2D_2{\partial^2\over\partial\alpha\partial
\alpha^*})W(\alpha,\alpha^*,t).\eqno(3.1)$$
When we substitute the $P$ representation function $P(\alpha,\alpha^*,
t)$ for $W(\alpha,\alpha^*,t)$ and the coefficients $D_1+\mu$ for $D_1
, D_2-\lambda$ for $D_2$ in the above equation, we get the
Fokker-Planck equation for the coherent representation \cite{i2}.

The Fokker-Planck equation for the $P$ representation, subject to the
initial condition
$$P(\alpha,\alpha^*,0)=\delta(\alpha-\alpha_0)\delta(\alpha^*-\alpha^
*_0),\eqno(3.2)$$
where $\alpha_0$ is the initial value of $\alpha$ can be solved
\cite{ga,ha,r} and the solution (Green function) for the $P$
representation is found to be
$$P(\alpha,\alpha^*,t)={2\over \pi\sqrt{{\rm det}\sigma(t)}}\exp\{-{1
\over{\rm det} \sigma(t)}[\sigma_{22}(\alpha-\bar\alpha_0)^2+\sigma_
{11}(\alpha^*-\bar\alpha_0^*)^2-2\sigma_{12}\vert\alpha-\bar\alpha_0
\vert^2\},\eqno(3.3)$$
where
$$\sigma_{ij}(t)=\sum_{s,r=1,2}[\delta_{is}\delta_{jr}-b_{is}(t)b_{jr}
(t)]\sigma_{sr}(\infty).\eqno(3.4)$$
The function $\bar\alpha_0$ and its complex conjugate, which are still
functions of time, are given by
$$\bar\alpha_0=b_{11}(t)\alpha_0+b_{12}(t)\alpha_0^*.\eqno(3.5)$$
The functions $b_{ij}$ obey the equations
$$\dot b_{is}=\sum_{j=1,2}c_{ij}b_{js}\eqno(3.6)$$
with the initial conditions $b_{js}(0)=\delta_{js}$ and $\sigma
(\infty)$ is
determined by
$$C\sigma(\infty)+\sigma(\infty)C^{\rm T}=Q^P,\eqno(3.7)$$
where
$$C=\left(\matrix{\lambda+i\omega&-\mu\cr
-\mu&\lambda-i\omega\cr}\right),~~~
Q^P=\left(\matrix{D_1+\mu&D_2-\lambda\cr
D_2-\lambda&D_1^*+\mu\cr}\right).\eqno(3.8)$$
We get
$$b_{11}=b_{22}^*=e^{-\lambda t}(\cos\Omega t-i{\omega\over\Omega}\sin
\Omega t),~~b_{12}=b_{21}={\mu\over\Omega}e^{-\lambda t}\sin\Omega t,
\eqno(3.9)$$
with $\Omega^2\equiv\omega^2-\mu^2.$ While the functions $\sigma_{11},
\sigma_{22}$ and $\bar\alpha_0$ are complex with $\sigma_{11}=\sigma_
{22}^*$, the functions ${\rm det}\sigma(t)$ and $\sigma_{12}$ are
real.

The solution of the Fokker-Planck equation has been written down
providing the diffusion matrix $Q^P$ is positive definite. However,
the diffusion matrix in the Glauber $P$ representation is not, in
general, positive definite. If the $P$ distribution does not exist as
a well-behaved function, the so-called generalized $P$ distributions
can be taken that are well-behaved, normal ordering functions
\cite{dg}.

In the coherent representation \cite{gl,h5} the density operator
$\rho(t)$
is expressed by
$$\rho(t)=\int P(\alpha,\alpha^*,t)\vert\alpha><\alpha\vert d^2\alpha,
\eqno(3.10)$$
where $d^2\alpha=d({\rm Re}\alpha)d({\rm Im}\alpha)$ and $\vert\alpha>
$ is the coherent state. The matrix element of $\rho(t)$ in the $n$
quantum number representation is obtained by multiplying (3.10) on the
left by $<m\vert$ and on the right by $\vert n>.$ By making use of the
well-known relation
$$\vert\alpha>=\exp(-{1\over 2}\vert\alpha\vert^2)\sum_{n=0}^\infty{
\alpha^n\over\sqrt{n!}}\vert n>,\eqno(3.11)$$
we get
$$<m\vert\rho(t)\vert n>={1\over\sqrt{m!n!}}\int\alpha^n\alpha^{*m}P(
\alpha,\alpha^*,t)\exp(-\vert\alpha\vert^2)d^2\alpha.\eqno(3.12)$$
Upon introducing the explicit form (3.3) for $P(\alpha,\alpha^*,t)$
into (3.12), we obtain the desired density matrix for the initial
coherent state. However, due to the powers of complex variables
$\alpha$ and $\alpha^*$ in the integrand, the practical evaluation of
the integral in (3.12) is not an easy task. Instead, we use the method
of generating function \cite{j} which allows us to transform (3.12)
into a
multiple-differential form. When we define a generating function
$F(x,y,t)$ by the integral
$$F(x,y,t)=\int P(\alpha,\alpha^*,t)\exp(-\vert\alpha\vert^2+x\alpha+y
\alpha^*)d^2\alpha,\eqno(3.13)$$
we see that the density matrix is related to the generating function
by
$$<m\vert\rho(t)\vert n>={1\over\sqrt{m!n!}}({\partial\over\partial x}
)^m({\partial\over\partial y})^nF(x,y,t)\vert_{x=y=0}.\eqno(3.14)$$
Since the $P$ representation is in a Gaussian form, the right-hand
side of (3.13) can be evaluated analitically by making use of the
identity
$$\int\exp(-a\vert z\vert^2+bz+cz^*+ez^2+fz^{*2})d^2z={\pi\over\sqrt{
a^2-4ef}}\exp{abc+b^2f+c^2e\over a^2-4ef},\eqno(3.15)$$
which is convergent for ${\rm Re}~a>\vert e^*+f\vert$, while $b,c$ may
be arbitrary. We find:
$$F={2\over \sqrt {\vert A\vert}}\exp\{xy-{1\over A}[\sigma_{11}(x-
\bar\alpha_0^*)^2+\sigma_{22}(y-\bar\alpha_0)^2-2(\sigma_{12}+2)(x-
\bar\alpha_0^*)(y-\bar\alpha_0)]\},\eqno(3.16)$$
where $A\equiv d-4(\sigma_{12}+1), ~~d\equiv {\rm det}\sigma=\sigma_
{11}\sigma_{22}-\sigma_{12}^2.$ A formula for the density matrix can
be written down by applying the relation (3.14) to the generating
function (3.16). We get
$$<m\vert\rho(t)\vert n>=2{\sqrt{m!n!}\over \sqrt {\vert A\vert}}\exp
\{[-{1\over A}[\sigma_{22}\bar\alpha_0^2+\sigma_{11}\bar\alpha_0^{*2}-
2(\sigma_{12}+2)\vert\bar\alpha_0\vert^2]\}$$
$$\times\sum_{n_1,n_2,n_3=0}{(-1)^{n_1+n_2}2^{m+n-2(n_1+n_2+n_3)}\over
n_1!n_2!n_3!(m-2n_1-n_3)!(n-2n_2-n_3)!}E,\eqno(3.17)$$
where
$$E={\sigma_{11}^{n_1}\sigma_{22}^{n_2}(d-2\sigma_{12})^{n_3}[\sigma_
{11}\bar\alpha_0^*-(\sigma_{12}+2)\bar\alpha_0]^{m-2n_1-n_3}[\sigma_
{22}\bar\alpha_0-2(\sigma_{12}+2)\bar\alpha_0^*]^{n-2n_2-n_3}\over A^
{m+n-(n_1+n_2+n_3)}}.$$
The expression (3.17) is the density matrix corresponding to the
initial coherent state. At time $t=0,$ the functions $\sigma_{11},$
$\sigma_{22}$ and $\sigma_{12}$ vanish and $\bar\alpha_0$ goes to
$\alpha_0.$ In this case the density matrix reduces to
$$<m\vert\rho(0)\vert n>={1\over\sqrt{m!n!}}\alpha_0^n\alpha_0^{*m}
\exp(-\vert\alpha_0\vert^2),\eqno(3.18) $$
which is the initial Glauber packet. For the diagonal case the initial
density matrix becomes the Poisson distribution. At infinity of time,
the density matrix (3.17) goes to the Bose-Einstein distribution
$$<m\vert\rho(\infty)\vert n>={<n>^m\over(1+<n>)^{m+1}}\delta_{mn}.
\eqno(3.19)$$

\subsection{Wigner distribution function}

The Fokker-Planck equation (3.1) can also be written in terms of real
coordinates $x_1$ and $x_2$ (or the averaged position and momentum
coordinates of the harmonic oscillator) defined by $\alpha=x_1+ix_2,
~\alpha^*=x_1-ix_2,$ as follows:
$${\partial W\over\partial t}=\sum_{i,j=1,2}A_{ij}{\partial\over
\partial x_i}(x_jW)+{1\over 2}\sum_{i,j=1,2}Q^W_{ij}{\partial^2\over
\partial x_i\partial x_j}W,\eqno(3.20)$$
where
$$A=\left(\matrix{\lambda-\mu&-\omega\cr
\omega&\lambda+\mu\cr}\right),~~~
Q^W={1\over\hbar}\left(\matrix{m\omega D_{qq}&D_{pq}\cr
D_{pq}&D_{pp}/m\omega\cr}\right).\eqno(3.21)$$
Since the drift coefficients are linear in the variables $x_1$ and
$x_2$ and the diffusion coefficients are constant with respect to
$x_1$ and $x_2,$ (3.20) describes an Ornstein-Uhlenbeck process
\cite{w,u}. Following the method developed by Wang and Uhlenbeck
\cite{w}, we
shall solve this Fokker-Planck equation, subject to either the
wave-packet type or the $\delta$-function type of initial conditions.
By changing the variables $x_1$ and $x_2$ of (3.20) via the relations
$$z_1=ax_1+bx_2,~~ z_2=cx_1+dx_2,\eqno(3.22)$$
(3.20) is transformed into the standard partial differential equation
\cite{w} expressed as
$${\partial W(z_1,z_2,t)\over\partial t}=(-\sum_{i=1,2}\nu_i{\partial
\over\partial z_i}z_i+{1\over 2}\sum_{i,j=1,2}D_{ij}{\partial^2\over
\partial z_i\partial z_j})W(z_1,z_2,t).\eqno(3.23)$$
In deriving this equation we have put $a=c^*=(\mu-i\Omega)/\omega,
b=d=1, \nu_1=\nu_2^*\equiv-\lambda-i\Omega$ and
$$D_{11}=D_{22}^*\equiv{1\over\hbar\omega}[(\mu-i\Omega)^2mD_{qq}+2(
\mu-i\Omega)D_{pq}+{D_{pp}\over m}],$$
$$D_{12}=D_{21}\equiv{1\over\hbar}(m\omega D_{qq}+{2\mu\over\omega}D_
{pq}+{D_{pp}\over m\omega}).\eqno(3.24)$$

1) When the Fokker-Planck equation for the coherent state
representation is subject to the initial condition $\delta(\alpha-
\alpha_0)\delta(\alpha^*-\alpha^*_0),$ then the use of the relation
between the Wigner distribution function and $P$ representation
$$W(\alpha,\alpha^*,t)={2\over\pi}\int P(\beta,\beta^*,t)\exp(-2\vert
\alpha-\beta\vert^2)d^2\beta\eqno(3.25)$$
leads to a Gaussian form for the initial Wigner function. If this
Wigner function is expressed in terms of $x_{10}$ and $x_{20}$ -- the
initial values of $x_1$ and $x_2$ at $t=0,$ respectively, then we get
the expression which corresponds to the initial condition of a wave
packet:
$$W_w(x_1,x_2,0)=
{1\over 2\hbar}W_w(\alpha,\alpha^*,0)={1\over\pi\hbar}\exp(-2\vert\alpha-
\alpha_0\vert^2)={1\over\pi\hbar}\exp\{-2[(x_1-x_{10})^2+(x_2-x_{20})^2]\}.
\eqno(3.26)$$
Accordingly, we now look for the solution of the Fokker-Planck
equation (3.20) subject to (3.26). By changing the variables $x_1$ and
$x_2$ into $z_1$ and $z_2, (z_1=z^*_2\equiv z)$, this initial
condition is seen to be transformed into
$$W_w(z,z^*,0)={1\over\pi\hbar}\exp\{{2\omega^2\over \Omega^2}[q(z-z_0)^2+q
^*(z^*-z^*_0)^2-\vert z-z_0\vert^2]\},\eqno(3.27)$$
where $z_0$ is the initial value of $z$ and $q=\mu(\mu+i\Omega)/2
\omega^2.$ The solution of (3.23) subject to the initial condition
(3.27) is found to be
$$W_w(z,z^*,t)={\Omega\over \pi\hbar\omega\sqrt{|B_w|}}\exp\{-{1
\over 2B_w}[g_2(z-z_0e^{\nu_1t})^2+g_1(z^*-z_0^*e^{\nu_2t})^2-g_3\vert
z-z_0e^{\nu_1t}\vert ^2]\},\eqno(3.28)$$
where
$$B_w=g_1g_2-{1\over 4}g_3^2, g_1=g_2^*=q^*e^{2\nu_1 t}+{D_{11}\over 2
\nu_1}(e^{2\nu_1t}-1), g_3=e^{-2\lambda t}+{D_{12}\over\lambda}(1-e^{-
2\lambda t}).\eqno(3.29)$$
In terms of real variables $x_1$ and $x_2$ we have:
$$W_w(x_1,x_2,t)={\Omega\over \pi\hbar\omega\sqrt{|B_w|}}\exp\{-{1\over 2B_w}
[\phi_w(x_1-\bar x_1)^2+\psi_w(x_2-\bar x_2)^2+\chi_w(x_1-\bar x_1)
(x_2-\bar x_2)]\},\eqno(3.30)$$
where
$$\phi_w=g_1a^{*2}+g_2a^2-g_3,~\psi_w=g_1+g_2-g_3,~\chi_w=2(g_1a^*+g_
2a)-g_3(a+a^*).\eqno(3.31)$$
The functions $\bar x_1$ and $\bar x_2$, which are also oscillating
functions, are given by
$$\bar x_1=e^{-\lambda t}[x_{10}(\cos\Omega t+{\mu\over\Omega}\sin
\Omega t)+x_{20}{\omega\over\Omega}\sin\Omega t],$$
$$\bar x_2=e^{-\lambda t}[x_{20}(\cos\Omega t-{\mu\over\Omega}\sin
\Omega t)-x_{10}{\omega\over\Omega}\sin\Omega t].\eqno(3.32)$$

2) If the Fokker-Planck equation (3.23) is subject to the
$\delta$-function type of initial condition, the Wigner distribution
function is given by
$$W(z,z^*,t)={\Omega\over\pi\hbar\omega\sqrt{\vert B\vert}}\exp\{-{1\over
B}[f_2(z-z_0e^{\nu_1t})^2+f_1(z^*-z_0^*e^{\nu_2t})^2-2f_3\vert z-z_0
e^{\nu_1t}\vert^2]\},\eqno(3.33)$$
where
$$B=f_1f_2-f_3^2,~~f_1=f_2^*={D_{11}\over \nu_1}(e^{2\nu_1t}-1),~~
f_3={D_{12}\over \lambda}(1-e^{-2\lambda t}).\eqno(3.34)$$
In terms of real variables $x_1$ and $x_2$ we have:
$$W(x_1,x_2,t)={\Omega\over\pi\hbar\omega\sqrt{\vert B\vert}}\exp\{-{1\over B}
[\phi_d(x_1-\bar x_1)^2+\psi_d(x_2-\bar x_2)^2+\chi_d(x_1-\bar x_1)
(x_2-\bar x_2)]\},\eqno(3.35)$$
where
$$\phi_d=f_1a^{*2}+f_2a^2-2f_3,~\psi_d=f_1+f_2-2f_3,~\chi_d=2[f_1a^*+f_
2a-f_3(a+a^*)].\eqno(3.36)$$
So, one gets a 2-dimensional Gaussian distribution with the average
values $\bar x_1$ and $\bar x_2$ and the variances $\phi_d,$$\psi_d$
and $\chi_d.$

When time $t\to\infty,$ $\bar x_1$ and $\bar x_2$ vanish and
we obtain the steady state solution:
$$W(x_1,x_2)={1\over 2\pi\sqrt{{\rm det}\sigma^W(\infty)}}\exp[-{1
\over 2}\sum_{i,j=1,2}(\sigma^W)^{-1}_{ij}(\infty)x_ix_j].
\eqno(3.37)$$
The stationary covariance matrix $\sigma^W(\infty)$ can be determined
from the algebraic equation
$$A\sigma^W(\infty)+\sigma^W(\infty)A^{\rm T}=Q^W.\eqno(3.38)$$
We obtain:
$$\sigma_{11}^W(\infty)={(2\lambda(\lambda+\mu)+\omega^2)Q_{11}^W+
\omega^2Q_{22}^W+2\omega(\lambda+\mu)Q_{12}^W\over 4\lambda(\lambda^2+
\omega^2-\mu^2)},$$
$$\sigma_{22}^W(\infty)={\omega^2Q_{11}^W+(2\lambda(\lambda-\mu)+
\omega^2)Q_{22}^W-2\omega(\lambda-\mu)Q_{12}^W\over 4\lambda(\lambda^2
+\omega^2-\mu^2)},\eqno(3.39)$$
$$\sigma_{12}^W(\infty)={-\omega(\lambda+\mu)Q_{11}^W+\omega(\lambda-
\mu)Q_{22}^W+2(\lambda^2-\mu^2)Q_{12}^W\over 4\lambda(\lambda^2+\omega
^2-\mu^2)}.$$

\section{Conclusions}

Recently we assist to a revival of interest in quantum Brownian motion
as a paradigm of quantum open systems. There are many motivations. The
possibility of preparing systems in macroscopic quantum states led to
the problems of dissipation in tunneling and of loss of quantum
coherence (decoherence). These problems are intimately related to the
issue of quantum-to-classical transition. All of them point the
necessity of a better understanding of open quantum systems and all
requires the extension of the model of quantum Brownian motion. The
Lindblad theory provides a selfconsistent treatment of damping as a
possible extension of quantum mechanics to open systems. In the
present paper we have studied the one-dimensional harmonic oscillator
with dissipation within the framework of this theory. We have carried
out a calculation of the expectation values of various dynamical
operators involved in the master equation, especially the first two
moments and the density matrix. Generally, the time evolution of the
density matrix as well as the expectation values of dynamical
operators are characterized by complex functions with an oscillating
element $\exp(\pm i\sqrt{\omega^2-\mu^2}t)$ multiplied by the damping
factor $\exp(-\lambda t).$ We deduced the density matrix from the
solution of the Fokker-Planck equation for the coherent state
representation, obtained from the master equation for the density
operator. For a thermal bath, when the asymptotic state is a Gibbs
state, a Bose-Einstein distribution results as density matrix. The
density matrix can be used in various physical applications where a
Bosonic degree of freedom moving in a harmonic oscillator potential is
damped. For example, one needs to determine nondiagonal transition
elements of the density matrix, if an oscillator is perturbed by a
weak electromagnetic field in addition to its coupling to a heat bath.
From the master equation of the damped quantum oscillator we have also
derived the corresponding Fokker-Planck equation in the Wigner $W$
representation. The obtained equation describes an Ornstein-Uhlenbeck
process. By using the Wang-Uhlenbeck method we have solved this
equation for the Wigner function, subject to either the Gaussian type
or the $\delta$-function type of initial conditions and showed that
the Wigner functions are two-dimensional Gaussians with different
widths. In a forthcoming paper, by using this Wigner function, it will
be discussed the entropy of the damped harmonic oscillator.

\end{document}